\documentclass{emulateapj}
\usepackage{verbatim,graphicx,natbib, amsmath}
\shortauthors{Cowan \& Abbot}
\shorttitle{Super-Earths Need Not be Waterworlds}

\def \apjl {Astrophysical Journal Letters}
\def \apj {Astrophysical Journal}
\def \nat {Nature}
\def \jgr {Journal of Geophysical Research}
\def \grl {Geophysical Research Letters}
\def \width {90mm}
\def \widewidth {180mm}

\begin{document}

\title{Water Cycling Between Ocean and Mantle: Super-Earths Need Not be Waterworlds}
\author{Nicolas B. Cowan\altaffilmark{1} \& Dorian S. Abbot\altaffilmark{2}}

\altaffiltext{1}{Center for Interdisciplinary Exploration and Research in Astrophysics (CIERA), Department of Earth \& Planetary Sciences, Department of Physics \& Astronomy, Northwestern University, 2145 Sheridan Road, Evanston, IL, 60208, USA\\ n-cowan@northwestern.edu}
\altaffiltext{2}{Department of the Geophysical Sciences, University of Chicago, 5734 South Ellis Avenue, Chicago, IL, 60637, USA}

\begin{abstract}
Large terrestrial planets are expected to have muted topography and deep oceans, implying that most super-Earths should be entirely covered in water, so-called waterworlds. This is important because waterworlds lack a silicate weathering thermostat so their climate is predicted to be less stable than that of planets with exposed continents. In other words, the continuously habitable zone for waterworlds is much narrower than for Earth-like planets.  A planet's water is partitioned, however, between a surface reservoir, the ocean, and an interior reservoir, the mantle.  Plate tectonics transports water between these reservoirs on geological timescales. Degassing of melt at mid-ocean ridges and serpentinization of oceanic crust depend negatively and positively on seafloor pressure, respectively, providing a stabilizing feedback on long-term ocean volume. Motivated by Earth's approximately steady-state deep water cycle, we develop a two-box model of the hydrosphere and derive steady-state solutions to the water partitioning on terrestrial planets. Critically, hydrostatic seafloor pressure is proportional to surface gravity, so super-Earths with a deep water cycle will tend to store more water in the mantle.  We conclude that a tectonically active terrestrial planet \emph{of any mass} can maintain exposed continents if its water mass fraction is less than $\sim0.2$\%, dramatically increasing the odds that super-Earths are habitable. The greatest source of uncertainty in our study is Earth's current mantle water inventory: the greater its value, the more robust planets are to inundation. Lastly, we discuss how future missions can test our hypothesis by mapping the oceans and continents of massive terrestrial planets.   
\end{abstract}

\section{Feedback or Luck?}
The stochastic delivery of water combined with its low density relative to rock leads to the generic expectation that many terrestrial planets should be entirely covered in water, so-called waterworlds \citep{Raymond_2004, Morbidelli_2012}.\footnote{\cite{matsui1986evolution} proposed that the mass of Earth's hydrosphere was set early-on by magma ocean buffering of a steam atmosphere, independent of the water content of Earth's building blocks. But this hypothesis does not explain the long-term stability of Earth's surface water reservoir.}  Massive terrestrial planets, ``super-Earths,'' are more likely to be waterworlds: a planet's mass scales faster than its surface area, so bigger planets ought to have deeper oceans, while their increased surface gravity produces shallow ocean basins \citep{Kite_2009, Abbot_2011}. 

Detrital zircons indicate that Earth has had both oceans and exposed continents for roughly 4.5~Gyrs \citep{wilde2001evidence}, while stratigraphy of sedimentary deposits suggests that the average height of continental surfaces above sea level ---freeboard--- has remained approximately constant since at least 2.5 Ga \citep{wise1974continental, eriksson1999sea}.  This is remarkable given the growth of continental crust over time \citep{Harrison_2009}, and the continuous two-way exchange of water between ocean and mantle \citep{McGovern_1989}.  

There are two classes of explanations for Earth's geologically stable surface character \citep{Kasting_1992}: blind luck in water delivery \citep{Raymond_2004} and ocean--mantle fluxes \citep{McGovern_1989}, or the existence of a stabilizing feedback \citep{Kasting_1992, Holm_1996, Abbot_2012}. 

The partitioning of water is a key factor in regulating planetary climate: surface water is the source of atmospheric water vapor, provides thermal inertia, and helps transport heat. Moreover, clement surface conditions on Earth are hypothesized to have been maintained over geological time by a silicate weathering feedback \citep{Walker_1981}. The silicate weathering thermostat requires exposed continents because their chemical weathering is strongly temperature-dependent.  Critically, the traditional conception of the habitable zone assumes a silicate weathering thermostat \citep{Kasting_1993}.  Barring the existence of a seafloor weathering feedback, waterworlds will not have a silicate weathering thermostat and should have a much narrower habitable zone \citep{Abbot_2012}.  Dry planets could have a wider habitable zone in principle \citep{Abe_2011}, but may lack the plate tectonics \citep{mian1990no} and physical erosion \citep{West_2005} necessary to maintain a silicate weathering thermostat.  

\subsection{Water Capacity of Earth's Mantle}\label{water_capacity}
The exact water inventory of Earth's interior is currently unknown, but is thought to be comparable to the surface reservoirs. The mantle contains water both in hydrous and nominally anhydrous minerals \citep{hirschmann2006water, Hirschmann_2012}. The Fe core may also contain primordial water \citep{abe2000water}, which we ignore in the present analysis. 

\cite{Inoue_2010} determined maximum water mass fractions of 0.7\% (olivine, shallower than 410~km), 3.3\% (wadsleyite, 410--520~km), 1.7\% (ringwoodite, 520--660~km), and 0.1\% (perovskite, below 660~km). Combining these estimates with the pressure-dependence of olivine water content from \cite{Hauri_2006} puts Earth's mantle water capacity at a dozen times the current surface reservoir. Measurements of electrical conductivity \citep{dai2009electrical} suggest that Earth's mantle contains 1--2 oceans worth of water at present-day, but other estimates have differed by a factor of a few in either direction \citep[e.g.,][]{huang2005water, smyth2006nominally, khan2012geophysical}. 

\subsection{The Deep Water Cycle}
There is a two-way flux of water between ocean and mantle on Earth. Ocean crust forms at mid-ocean ridges by depressurization melting of mantle, releasing volatiles into the ocean. Plate tectonics then drives the ocean crust toward subduction zones. Hydrothermal alteration, principally serpentinization, produces ocean crust that is 5--10\% water by mass. During subduction, much of the water is baked out of the subducting slab, producing explosive volcanism like Mount St.\ Helens.  Some of the water, however, is subducted deep into the mantle, closing the loop \citep[for a review of the relevant geochemistry, see][]{Arndt_2013}. 

The current H$_2$O degassing at mid-ocean ridges is approximately $2\times10^{11}$~kg/yr \citep{Hirschmann_2012}, while the regassing flux at subduction zones is estimated to be 0.7--2.9$\times10^{12}$~kg/yr \citep{Jarrard_2003, Schmidt_1998}. Given the order-of-magnitude agreement of these figures, and the large associated uncertainties, Earth's current ocean volume is typically assumed to be in a steady state \citep{McGovern_1989}. Moreover, realistic parameterizations of mantle convection predict water fluxes of $10^{11}$--$10^{13}$~kg/yr in Earth's geological past, implying ocean-cycling times of $10^8$~yrs \citep{McGovern_1989}. A significant flux imbalance would have long ago desiccated or submerged the planetary surface.

\subsection{Previous Work}
A planet entirely covered in water may develop exposed continents by losing water to space, hydrating the crust and mantle, or reshaping continents and ocean basins. For example, an ocean-covered planet with a hot stratosphere could lose water to space until continents are exposed. The resulting vigorous silicate weathering might draw down sufficient CO$_2$ to cool the planet out of the moist greenhouse state, leaving a partially water-covered planet \citep{Abbot_2012}. Detailed atmospheric simulations of such hot planets, however, indicate that the tendency of carbon dioxide to cool the stratosphere impedes the loss of water \citep{Wordsworth_2013}.  

In an alternate hypothesis, the Rayleigh number of convecting water in hydrothermal systems at mid-ocean ridges exhibits a sharp peak when water becomes supercritical, which occurs at nearly the pressures and temperatures in hydrothermal systems today \citep{Kasting_1992}.  If Earth's water started in the mantle and gradually degassed to the surface, then this pressure-dependence could explain why the oceans stabilized at their current depths.  A planet starting with very deep oceans \citep{Korenaga_2008}, however, would not be able to effectively subduct water and would remain a waterworld.    

In order to solve the problem of regassing water into the mantle, \cite{Holm_1996} suggested that the reduced efficiency of hydrothermal heat exchange beneath deep oceans would result in higher mantle temperatures, greater plate velocities and therefore efficient subduction of hydrated crust.  This mechanism may not work quantitatively, however, because hydrothermal heat transfer only accounts for 30--50\% of heat flux through oceanic crust \citep{stein1995heat}. There are also qualitative problems with this regassing argument: first of all, the same logic should apply to planets with very shallow oceans, potentially negating the supercritical circulation hypothesis.   Moreover, hydrating the mantle lowers its viscosity, further increasing plate velocities and making this mechanism a destabilizing feedback. Lastly, although rapid subduction may aid regassing, a hotter mantle almost certainly hinders it \citep{Bounama_2001}.  In short, it is not clear that reduced heat transport through the ocean crust leads to net regassing of water. 

Whatever geophysical processes govern Earth's deep water cycle presumably operate on other terrestrial planets with plate tectonics. The partitioning of water on Earth is not only an outstanding problem in geophysics \citep{Fyfe_1994, Langmuir_2012}, but one with important implications for the surface conditions and climate of terrestrial exoplanets.

\section{Hydrosphere Model}
\begin{figure}
\vspace{-2.0cm}
\begin{center}
      \includegraphics[angle=270, width=\width]{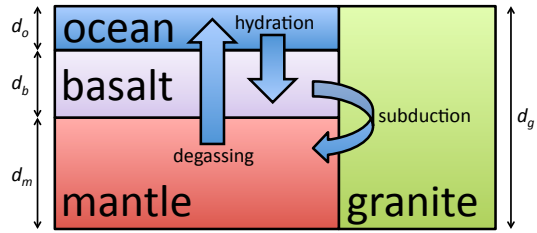}
      \end{center}
\vspace{-2.0cm}
    \caption{A schematic of our hydrosphere model.  Degassing of melt decreases with seafloor pressure, while hydration of ocean crust increases with seafloor pressure, providing a potential stabilizing feedback. \label{flux_schematic}}
\end{figure}

As shown in Figure~\ref{flux_schematic}, we develop a two-box model of the deep water cycle for a terrestrial planet with plate tectonics: the two water reservoirs are the ocean and mantle, while the basalt and granite are important for the transport of water and for setting the depth of ocean basins.  A two-box model is appropriate for gross estimates of water partitioning over billions of years \citep{McGovern_1989, Ueta_2013}, but may be too simple to simulate small changes in ocean volume on shorter timescales \citep{Parai_2012}.

We assume that the total hydrosphere mass, $W$, is constant, i.e., we neglect water loss from the top of the atmosphere. Isotopic evidence indicates that Earth's hydrosphere was at most 26\% more massive in the Archean \citep{Pope_2012}.  

\subsection{Water Fluxes}
The principle dependent variable in our model is the bulk water mass fraction of the mantle, $x$. We assume plate tectonics but our results should be largely insensitive to plate velocity: degassing and regassing are proportional to the spreading and subduction rates, respectively, which are equal in steady-state. We therefore consider the change in mantle water \emph{per sea-floor overturning}, $x' \equiv dx/d\tau$, where the non-dimensional time, $\tau$, is related to the area of oceanic crust, $A_b$, spreading rate, $S$, and the length of mid-ocean ridges, $L$, via $\tau = tLS/A_b$.

The net change of mantle water per sea-floor overturning is:
\begin{equation}\label{water_partitioning}
x'  = \frac{A_b}{M_m}(w_\downarrow - w_\uparrow),
\end{equation}
where $M_m$ is the mantle mass, $w_\downarrow$ is the water content of subducted crust and sediments, and $w_\uparrow$ is the water degassed by the formation of ocean crust at MORs (both have units of kg/m$^2$). This equation is similar to that solved by previous researchers \citep{McGovern_1989, Ueta_2013}, except that we include the first-order effects of seafloor pressure, as described below. 

The subducted water is
\begin{equation}
w_\downarrow = x_h \rho_{b} d_h(P) \chi,
\end{equation}
where $x_h=0.05$ is the mass fraction of water in the hydrated crust \citep{McGovern_1989}, $\rho_b=3.0\times10^3$~kg/m$^3$ is the density of basalt, $d_h$ is the depth of hydration in the crust (which depends on the seafloor pressure, $P$), and $\chi=0.23$ is the fraction of volatiles subducted deep into the mantle rather than outgassed via arc and back-arc volcanism \citep{rupke2004serpentine}.  The subduction efficiency, $\chi$, is highly uncertain but we perform a sensitivity analysis below to quantify how it and other critical parameters likely affect our results (\S\ref{sensitivity}).

The degassed water is
\begin{equation}
w_\uparrow = x \rho_m d_{\rm melt} f_{d}(P),
\end{equation}
where $\rho_m=3.3\times10^3$~kg/m$^3$ is the density of the upper mantle, $d_{\rm melt}=6\times10^4$~m is the depth of the MOR melting regime, and $f_{d}$ is the fraction of water in the melt that degasses rather than remaining in the ocean crust as it solidifies (model variables and parameters are listed in Table~\ref{parameters}). 

\begin{deluxetable}{lll}
\tabletypesize{\scriptsize}
\tablecaption{Model Variables \& Parameters \label{parameters}}
\tablewidth{0pt}
\tablehead{
\colhead{Name} & \colhead{Symbol} & \colhead{Value}}
\startdata
planetary water mass fraction$^{a}$ & $\omega$ & $\omega_\oplus = 6.2\times10^{-4}$\\
normalized gravity$^{a}$ & $\tilde{g}$ & $\tilde{g}_\oplus = 1$\\
mantle water mass fraction$^{a}$ & $x$ & $x_\oplus=5.8\times10^{-4}$\\
density of water & $\rho_w$ & $1.0\times10^3$ kg/m$^3$\\
density of granite & $\rho_g$ & $2.9\times10^3$ kg/m$^3$\\
density of basalt & $\rho_b$ & $3.0\times10^3$ kg/m$^3$\\
density of mantle & $\rho_m$ & $3.3\times10^3$ kg/m$^3$\\
thickness of oceanic crust & $d_b$ & $6\times10^3$ m\\
depth of melt & $d_{\rm melt}$ & $60\times 10^{3}$~m\\
hydrated crust water fraction & $x_h$ & 0.05\\
subduction efficiency & $\chi$ & 0.23\\
Earth degassing efficiency & $f_{d\oplus}$ & 0.9\\
Earth hydration depth & $d_{h\oplus}$ &  $3\times10^3$ m\\
Earth seafloor pressure & $P_\oplus$ & $4\times10^7$~Pa\\
max.\ thickness of cont.\ crust & $d_g^{\rm max}$ & $70\times10^3~\tilde{g}^{-1}$ m\\
ocean basin covering fraction & $f_b$ & 0.9\\
mantle mass fraction & $f_m$ & 0.68\\
ocean mass fraction of Earth & $\omega_o$ & $2.3\times10^{-4}$\\
normalized ocean basin area & $\tilde{f}_b$ & 1.3\\ 
seafloor pressure dependence$^b$ & $\phi$ & $2$\\
Earth mantle water content$^{b}$ & $x_\oplus$ & $5.8\times10^{-4}$\\
max mantle water fraction$^b$ & $x_{\rm max}$ & $7\times10^{-3}$
\enddata
\tablecomments{$^{a}$These are the variables in our model; we list here their nominal values for Earth. $^b$These are critical parameters whose sensitivity we test in \S\ref{sensitivity}; we list their nominal values here.}
\end{deluxetable}

\subsection{Degassing}
The fraction of water in the melt that is degassed depends on hydrostatic pressure at the seafloor \citep{Papale_1997, Kite_2009}. We parametrize this stabilizing feedback as:
\begin{equation}\label{f_d}
f_{d}(P) = \min\left[f_{d\oplus} \left(\frac{P}{P_\oplus}\right)^{-\mu} , \hspace{0.25cm}1\right],
\end{equation}
where $P$ is the pressure at the bottom of the ocean, $P_\oplus=4\times10^7$~Pa is its current value on Earth, $\mu>0$ quantifies the pressure-dependence of melt degassing, and $f_{d\oplus}=0.9$ is the nominal degree of melt degassing on Earth today. The piecewise definition ensures that the degassed water does not exceed that present in the melt.

\subsection{Regassing}
The depth of serpentinization may also depend on the pressure at the bottom of the ocean:
\begin{equation}\label{d_h}
d_h(P) = \min\left[d_{h\oplus} \left(\frac{P}{P_\oplus}\right)^\sigma ,\hspace{0.25cm}d_b\right],
\end{equation}
where $\sigma$ quantifies the pressure-dependence, $d_{h\oplus}=3\times10^3$~m is the nominal value of the hydration depth on Earth \citep{McGovern_1989}, and $d_b=6\times10^3$~m is the thickness of basaltic ocean crust. Hydration of the lithospheric mantle below the oceanic crust is functionally identical to the hydration of oceanic crust \citep{rupke2004serpentine}, but we conservatively adopt the constraint $d_h \le d_b$.

It has been argued that the depth of hydration depends on the Rayleigh number of water in hydrothermal systems \citep{Kasting_1992}. If, as they assumed, water is gradually degassed from the mantle, this amounts to a large positive $\sigma$; if water begins at the planetary surface such a mechanism dictates a large negative $\sigma$.

\subsection{Hypsometry \& Isostacy}\label{isostacy_section}
We treat the planetary elevation distribution (``hypsometry'') as two $\delta$-functions: one for ocean crust and another for continental crust.  This is a good approximation of Earth's strongly bimodal hypsometry \citep{Rowley_2013}, which is presumably typical of any terrestrial planet with plate tectonics \citep{Stoddard_2012}. Earth's oceanic crust exhibits a clear age-depth relation: older crust is denser and sinks deeper into the mantle \citep{parsons1977analysis}.  A larger ocean basin will therefore have a greater mean depth.  The globally averaged ocean depth, however, should remain roughly constant barring secular changes in spreading rates, which are beyond the scope of our zeroth-order treatement.  

We assume that the modal continental height is level with the ocean surface, which is natural because of the competing effects of erosion and deposition \citep{Korenaga_2008, Rowley_2013}. There is a maximum thickness that continents can achieve, however, beyond which a continent flows under its own weight \citep{rey2006lithospheric}. The Himalayan plateau on Earth has a crustal thickness of 70~km and appears to be near this limit \citep{england1982thin}.  We conservatively adopt a maximum continental thickness of 
\begin{equation}
d_g^{\rm max} = 70~\textrm{km}~\left(\frac{g}{g_\oplus}\right)^{-1}, 
\end{equation}
where $g$ is the surface gravity of the planet, $g_\oplus$ is that of Earth, and we have adopted the gravity-dependence of \cite{Kite_2009}. Using the mass-radius relation of \cite{Valencia_2007b}, a gravity of $3g_\oplus$ corresponds to a $10M_\oplus$ super-Earth.

The thickness of granitic continents, $d_g$, is related to the thickness of the other layers by:
\begin{equation}\label{total_thickness}
d_g = d_o + d_b + d_m,
\end{equation}
where $d_o$ is the depth of the ocean basin and $d_m$ is the depth to which the continent root extends into the mantle (Figure~\ref{flux_schematic}). This parameterization assumes that the tops of continents are at sea level (zero freeboard).  This is merely shorthand for a freeboard that is much smaller than the depth of ocean basins.

Isostatic balance dictates that the pressure below each vertical column must be equal:
\begin{equation}\label{isostatic_balance}
\rho_w d_o + \rho_b d_b + \rho_m d_m = \rho_g d_g,
\end{equation}
where the $\rho_w=1.0\times10^3$~kg/m$^3$ and $\rho_g=2.9\times10^3$~kg/m$^3$ are the density of water and granite, respectively.

Combining \eqref{total_thickness} and \eqref{isostatic_balance} yields the following expression of isostatic balance:
\begin{equation}\label{isostacy_eqn}
d_o (\rho_m-\rho_w)+d_b (\rho_m-\rho_b) = d_g (\rho_m-\rho_g).
\end{equation}

Given the maximal crustal thickness, $d_g^{\rm max}$, one can derive the maximum depth of water-filled ocean basins:
\begin{equation}\label{max_water}
d_{o}^{\rm max} = \frac{d_g^{\rm max} (\rho_m-\rho_g) - d_b (\rho_m-\rho_b)}{\rho_m-\rho_w}.
\end{equation}
For our fiducial Earth-size parameters, $d_{o}^{\rm max} = 11.4$~km. Note that the thickness of oceanic crust is likely also inversely proportional to $g$ \citep{Sleep_2012}. We conservatively adopt $d_b = 6$~km regardless of planet mass, which produces shallower ocean basins for massive planets.

The densities of granite and basalt are nearly the same ($\rho_g \approx \rho_b$) and the maximum crustal thickness far exceeds the thickness of oceanic crust ($d_b \ll d_g^{\rm max}$), so we can approximate \eqref{max_water} as
\begin{equation}\label{approx_max_water}
d_{o}^{\rm max} \approx 11.4~\textrm{km}~\left(\frac{g}{g_\oplus}\right)^{-1}. 
\end{equation}

The maximum ocean volume that can be accommodated while maintaining exposed continents also depends on the ocean basin area, $A_b = f_b A$, where $A$ is the planetary area.  We adopt $f_b = 0.9$. In other words, 90\% of the planet is covered in water and the remaining 10\% is exposed continent.  This is sufficient dry land to maintain a silicate weathering thermostat \citep{Abbot_2012} and our results would not change dramatically if we instead adopted $f_b = 0.7$ (as on modern Earth) or $f_b = 1$ (the limiting case of a single infinitesimal island extending out of the ocean). 

Given these assumptions, the maximum volume of surface water that could be accommodated by an Earth-size planet is $5.2\times 10^{18}$~m$^3$, or a mass of $3.7M_o$.

\section{Steady-State Solutions}
We find steady state solutions to the water partitioning on a planet by setting the upward and downward water fluxes equal to each other,
\begin{equation}\label{steady_state}
x \rho_m d_{\rm melt} f_d(P) = x_h \rho_{b} d_h(P) \chi.
\end{equation}
The degassing efficiency, $f_d$, and hydration depth, $d_h$, depend on seafloor pressure, $P$, which intimately depends on ocean water depth, $d_w$: $P = g \rho_w d_w$.

Ocean depth can be expressed in terms of mantle water content,
\begin{equation}
d_w = \frac{W - x M_m}{A_b \rho_w},
\end{equation}
but it is instructive to factor out the size-dependent terms and define intensive quantities:
\begin{equation}
d_w =  \left(\frac{M}{A}\right) \frac{\omega - x f_m}{f_b \rho_w},
\end{equation}
where $\omega = W/M$ is the planetary water mass fraction and $f_m = M_m/M = 0.68$ is the planetary mantle mass fraction.  Note that the term in parentheses is proportional to gravity: $M/A \propto M/R_p^2 \propto g$. We may therefore write the normalized ocean depth as
\begin{equation}\label{ocean_depth}
\frac{d_w}{d_{w\oplus}} = \tilde{g}~\frac{\omega - x f_m}{\omega_o\tilde{f}_b},
\end{equation}
where $d_{w\oplus} = 4$~km is the average depth of Earth's oceans, $\tilde{g} = g/g_\oplus$ is the normalized planetary gravity, $\omega_o = M_o/M_\oplus = 2.3\times 10^{-4}$ is the fractional mass of Earth's surface water, and $\tilde{f}_b=f_b/f_{b\oplus} = 1.3$ is the ocean basin covering fraction of the planet divided by that of Earth.

The normalized seafloor pressure is therefore
\begin{equation}\label{pressure}
\frac{P}{P_\oplus}=\tilde{g}^2~\frac{\omega - x f_m}{\omega_o \tilde{f}_b}. 
\end{equation}

We substitute \eqref{pressure} into \eqref{steady_state} and solve for the steady-state mantle water fraction on the interval $x \in [0,\omega/f_m]$, where the upper-limit ensures that the mantle does not contain more water than the planet as a whole.  There is also a petrological upper limit to how much water the mantle can sequester, however.  For Earth, that limit appears to be $x_{\rm max} = 7\times10^{-3}$ (a dozen oceans, \S\ref{water_capacity}). In cases where $x>x_{\rm max}$, we set $x=x_{\rm max}$.

\begin{figure*}
\begin{center}
      \includegraphics[width=\widewidth]{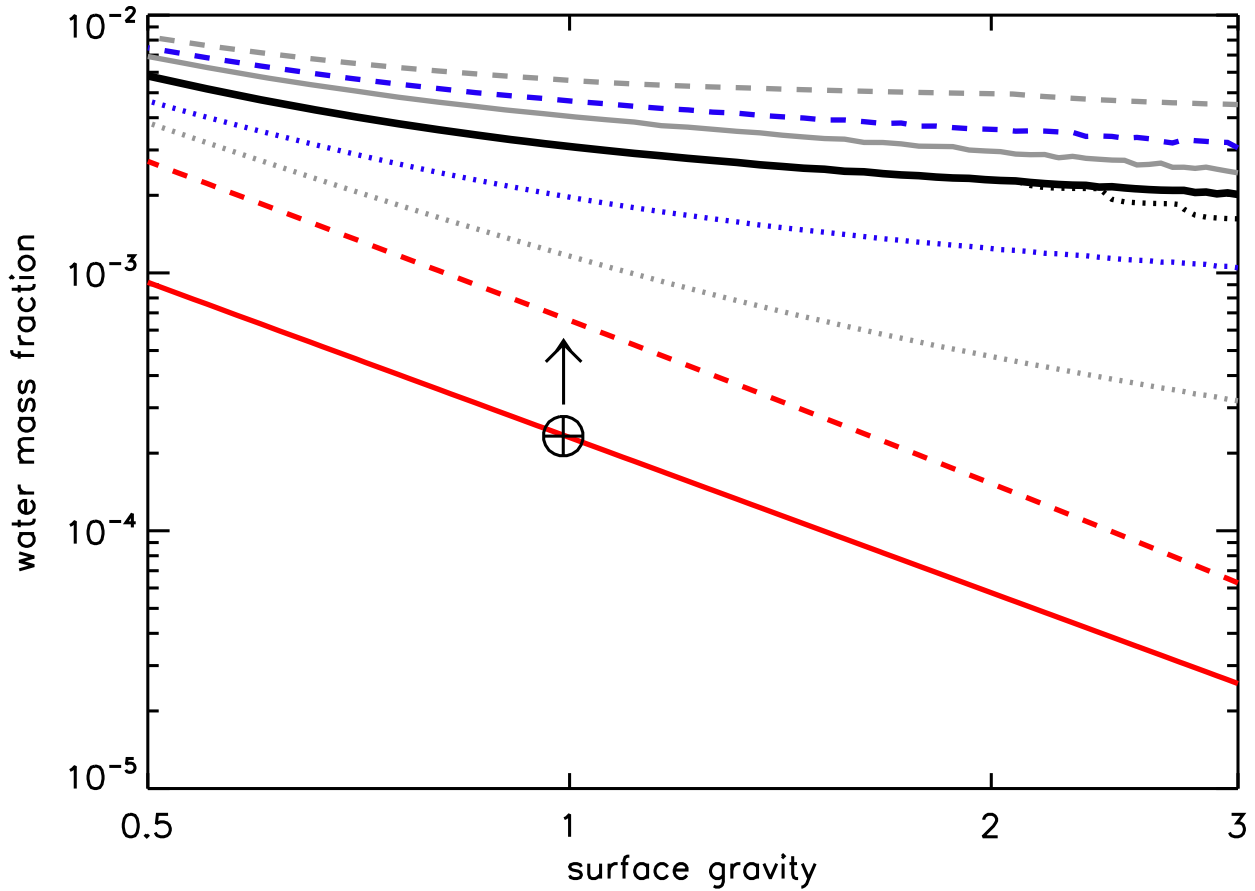}
      \end{center}
    \caption{Waterworld boundary as a function of water mass fraction, $\omega$, and normalized surface gravity, $\tilde{g} = g/g_\oplus$: planets in the upper-right corner are waterworlds, while those in the lower-left maintain exposed continents. The solid black line shows the waterworld boundary for our nominal parameters, including the negative feedbacks associated with seafloor pressure. The dotted black line shows the waterworld boundary if one assumes that water can only be stored in a mantle's transition zone.  The solid red line shows the waterworld boundary if one applies the $g^{-1}$ dependence to Earth's current hypsometry and presumes that all of a planet's water resides at the surface. The dashed red line accounts for the effects of erosion and isostatic adjustment (\S\ref{isostacy_section}) but not the deep-water cycle. The black symbol, $\oplus$, denotes Earth if one only considers its surface water reservoir; the arrow indicates Earth's probable location if one also accounts for water present in the mantle. The gray lines indicate the analytic waterworld boundary for nominal parameters (solid), as well as Earth mantle water content values of $x_\oplus = 5.8\times 10^{-3}$(dashed) and $x_\oplus = 5.8\times 10^{-5}$ (dotted). The blue lines show the waterworld boundary for seafloor pressure-dependencies of $\phi=3$ (dashed) and $\phi=1$ (dotted). A terrestrial planet with surface gravity $\tilde{g} = 3$ corresponds to a $10M_\oplus$ super-Earth \citep{Valencia_2007a}. \label{waterworld_boundary}}
\end{figure*}

Given the steady-state mantle water content, it is trivial to compute the depth of surface oceans using \eqref{ocean_depth}. The waterworld boundary is defined by equating surface water depth with maximal ocean basin depth, \eqref{max_water}. The solid black line in Figure~\ref{waterworld_boundary} shows the waterworld boundary for our fiducial parameters and pressure dependencies of $\mu=\sigma=1$.

It is useful to compare this waterworld boundary to the null hypthesis: ignoring isostatic adjustment and the deep water cycle.  In such a case, \eqref{ocean_depth} simplifies to $d_w = d_{w\oplus} \tilde{g} \omega/\omega_o$, which can be combined with the approximation \eqref{approx_max_water} to obtain the waterworld limit of $\omega = \omega_o \tilde{g}^{-2}$.  The null-hypothesis is indicated by the solid red line in Figure~\ref{waterworld_boundary}. By definition, Earth's surface reservoir (indicated by the $\oplus$) puts it right at the waterworld boundary under these assumptions.

If one accounts for the ability of isostacy and erosion to reshape continents and keep their heads above water, but still neglects the deep-water cycle, one obtains the dashed red line in Figure~\ref{waterworld_boundary}. 

\subsection{Analytic Approximation}
It is possible to obtain an analytic solution to \eqref{steady_state} if we ignore the piecewise nature of degassing, $f_d$, and regassing, $d_h$.  This analytic approximation has intuitive value so we develop it here.  In this case the $w_\uparrow = w_\downarrow$ can be written as:
\begin{equation}\label{analytic}
x \rho_m d_{\rm melt} f_{d\oplus} \left(\frac{P}{P_\oplus}\right)^{-\mu} = x_h \rho_{b} \chi d_{h\oplus} \left(\frac{P}{P_\oplus}\right)^\sigma,
\end{equation}
which can be compactly expressed as
\begin{equation}\label{compact}
\frac{x}{x_\oplus} = \left(\frac{P}{P_\oplus}\right)^\phi,
\end{equation}
where $\phi=\mu+\sigma$ is the sum of pressure dependencies for mid-ocean ridge melt degassing and serpentinization of oceanic crust, and Earth's bulk mantle water content is
\begin{equation}\label{x_Earth}
x_{\oplus} = \frac{x_h  \chi \rho_b d_{h\oplus}}{\rho_m d_{\rm melt} f_{d\oplus}}.
\end{equation}
Our fiducial parameter values yield $x_{\oplus}=5.8\times 10^{-4}$, or roughly an ocean's worth of water in Earth's mantle. 

Substituting \eqref{pressure} into \eqref{compact}, we obtain
\begin{equation}\label{anal_eq}
\frac{x}{x_\oplus} = \left(\tilde{g}^2~\frac{\omega - x f_m}{\omega_o \tilde{f}_b}\right)^\phi.
\end{equation}

The steady-state mantle water content allows us to estimate the ocean depth, via \eqref{ocean_depth}, which we then compare to the maximum ocean basin depth. The solid gray line in Figure~\ref{waterworld_boundary} shows the waterworld boundary for our nominal parameter values and $\phi=2$. The small difference between the numerical and analytic waterworld boundaries (solid black and gray lines, respectively) indicates that the piecewise definitions of $f_d(P)$ and $d_h(P)$ do not critically affect our results.

If $\phi<0$, the pressure feedback is destabilizing and there are two physical steady-state solutions: shallow and deep oceans, respectively. The current state of a planet will depend on initial conditions.  For $\phi \ge 0$, there is a single physical root to \eqref{anal_eq} and therefore a single steady-state solution. If there is no net pressure dependence to the deep water cycle ($\phi=0$), then all planets have the same mantle water content as Earth, $x\equiv x_\oplus$. 

In the presence of a modest stabilizing pressure-dependence ($\phi=1$), the steady-state mantle water fraction is
\begin{equation}\label{first_order}
x = \frac{\omega x_\oplus \tilde{g}^2}{\omega_o \tilde{f}_b + x_\oplus f_m \tilde{g}^2}.
\end{equation}
In the high-gravity limit, the mantle contains the entirety of the planet's water, $x = \omega/f_m$.  In practice this cannot occur for water-rich planets because of the finite water capacity of the mantle ($x\le x_{\rm max}$). Nonetheless, the higher gravity of super-Earths biases the deep water cycle in favor of mantle sequestration. Water inventory and mantle capacity are both proportional to planetary mass, producing a weak mass-dependence to the waterworld boundary. 

\section{Discussion}\label{discussion}
\subsection{Sensitivity Analysis}\label{sensitivity}
As noted in \S\ref{water_capacity}, the amount of water in Earth's mantle is poorly constrained. The gray broken lines in Figure~\ref{waterworld_boundary} show the analytic waterworld boundary if Earth's mantle water content, $x_\oplus$, is $10\times$ greater (dashed) and $10\times$ smaller (dotted) than our nominal value.  Note that varying $x_\oplus$ in this model is mathematically equivalent to varying any of the water flux parameters in \eqref{x_Earth}, many of which are uncertain (e.g., the subduction efficiency, $\chi$). Varying $x_\oplus$ by two orders of magnitude affects the waterworld boundary for a $10M_\oplus$ super-Earth by roughly one order of magnitude; this is the dominant uncertainty in our study.

The maximum water capacity of the mantle, $x_{\rm max}$ is not well known for high-mass terrestrial planets.  The water storage in Earth's mantle is thought to be concentrated in the transition zone (410--660~km depth). If massive planets can only sequester water in a thin transition zone, then mantle water capacity scales with planetary area rather than mass, or $x_{\rm max} \propto \tilde{g}^{-1}$. The dotted black line in Figure~\ref{waterworld_boundary} shows the minuscule effect of adopting this scaling.  

On the other hand, the mantle of a super-Earth should be primarily in the form of post-perovskite \citep{Valencia_2007a}, which may not have the same water capacity as Earth's dominant mantle rock, perovskite. We try setting $x_{\rm max} = 1$ and find that the waterworld boundary is unchanged if post-perovskite can hold unlimited water.

Finally, the blue lines in Figure~\ref{waterworld_boundary} show the waterworld boundary for pressure-dependences of $\phi=3$ (dashed) and $\phi=1$ (dotted). (In both cases we use $\mu = \sigma = \phi/2$ in the numerical model.) The precise strength of the seafloor pressure dependence of the deep water cycle affects the waterworld boundary by less than a factor of two. 

\subsection{Plate Tectonics}
Our model of the deep water cycle assumes plate tectonics, but it is currently unknown whether super-Earths are tectonically active. The increased heat flux of super-Earths should produce vigorous mantle convection \citep{Valencia_2007b, vanHeck_2011}, but the increased strength and buoyancy of crust on super-Earths may prohibit plate tectonics \citep{ONeill_2007, Kite_2009}.  Mantle convection may even exhibit hysteresis, such that planets with identical boundary conditions may or may not have plate tectonics, depending on initial conditions \citep{Lenardic_2012}.  

Alternatively, it has been suggested that surface water is more important than planetary mass for plate tectonics \citep{mian1990no, korenaga2010likelihood}. In a classic case of chicken-and-egg, \cite{van2008role} argue that a deep water cycle is a necessary, if not sufficient, condition for long-term plate tectonics.  

Continental crust formation is thought to be an inevitable by-product of plate tectonics in the presence of water \citep{rudnick1995making, Arndt_2013}, so super-Earths are likely to have large volumes of granitic crust. In fact, the large volume of continental crust combined with smaller maximal crustal thickness may lead to a planet entirely covered in continental crust. This does not greatly affect our results, provided that the planet remains tectonically active and that some regions have thicker crust than others (differences in crustal thickness scale as $g^{-1}$ for the same reason as ocean basin depth).

\subsection{Homogeneity of the Mantle}
The homogeneity of volatiles in Earth's lower mantle is questionable. The high $^3$He abundance of ocean islands has been attributed to a poorly-mixed lower mantle \citep{Kurz_1982}.  By extension, this hypothesis implies that Earth's lower mantle may hold much more water than what is inferred for the upper mantle.  \cite{Gonnermann_2009} argue, however, that the $^3$He abundance of the mantle is consistent with homogeneous composition. 

If the mantle is not well-mixed, then $x$ represents the water fraction of those regions sampled by the mid-ocean ridge melting and affected by subduction of oceanic crust. Indeed, the source of mid-ocean ridge basalts (MORB) appears to have maintained a constant water mass fraction, $x$, for billions of years, suggesting that subduction of hydrated oceanic crust is depositing water in the MORB source region \citep{hirschmann2006water}. 

\subsection{Observational Constraints}
Our model of the deep water cycle predicts that many super-Earths have exposed continents. It will eventually be possible to test this hypothesis by observationally determining the surface character of a large number of high-mass terrestrial planets \citep[see also discussion in][]{Abbot_2012}. 

Disk-integrated rotational multiband photometry of Earth, essentially the changing colors of a pale blue dot, encode information about continents, oceans and clouds \citep{Ford_2001}.  Such ``single-pixel'' observations have been used to construct coarse longitudinal color maps of Earth and Mars \citep{Cowan_2009,Cowan_2011c, Fujii_2011, hasinoff2011diffuse}, while simulations suggest that photometry spanning an entire planetary orbit could be used to construct a rough 2D color map \citep{Kawahara_2010, Kawahara_2011, Fujii_2012}. Finally, \cite{Cowan_2013a} showed that disk-integrated multiband photometry of a variegated planet can be inverted to obtain reflectance spectra of its dominant surface types, even if the number and colors of the surfaces are not known \emph{a priori}. 

The bottom line is that a 5--10~m space telescope equipped with a coronagraph or starshade could produce a coarse surface map of an Earth-analog at a distance of 10~pc \citep{Cowan_2009, Fujii_2012}.  Such low-resolution maps would be sufficient to identify the continents one expects on a tectonicaly active planet.  If most super-Earths exhibit the bimodal surface character of Earth, it will suggest that they experience plate tectonics and a deep water cycle. If, instead, large terrestrial planets were \emph{all} determined to be waterworlds, it would indicate that our hypothesis is wrong.  

\section{Conclusions}
It has long been suggested that super-Earths ought to be waterworlds \citep{Stapledon_1937, Kite_2009, Abbot_2011}. If one accounts for the first-order effects of gravity on ocean basin depth and water inventory, then a $10M_\oplus$ planet is not expected to have exposed continents unless it has a water mass fraction less than $3\times10^{-5}$, roughly ten times drier than Earth (solid red line in Figure~\ref{waterworld_boundary}). 

We have argued, as have others \citep{Kasting_1992, Holm_1996}, that the approximately steady-state water partitioning on Earth over geological time suggests a seafloor pressure feedback that regulates the degassing at mid-ocean ridges and/or the serpentinization and subsequent subduction of oceanic crust. Although ocean volume may change throughout a planet's history because of secular cooling, we have tackled the zeroth-order problem of steady-state solutions.

Notably, seafloor pressure is proportional to a planet's surface gravity. The enhanced gravity of super-Earths produces shallower ocean basins, but also leads to shallower oceans. The solid black line in Figure~\ref{waterworld_boundary} shows the waterworld boundary if one accounts for the pressure-dependence of the deep water cycle. 

The effects of isostacy, erosion, and deposition, combined with a pressure-dependent deep water cycle, make super-Earths $80\times$ less susceptible to inundation than they otherwise would be.  Our model predicts that tectonically active $10M_\oplus$ planets can maintain large exposed continents for water mass fractions less than $2\times10^{-3}$.  

Exoplanets with sufficiently high water content will be water-covered regardless of the mechanism discussed here, but such ``ocean planets'' may betray themselves by their lower density: a planet with 10\% water mass fraction will exhibit a transit depth 10\% greater than an equally-massive planet with Earth-like composition \citep{sotin2007mass}. Planets with 1\% water mass fraction, however, are almost certainly waterworlds but may have a bulk density indistinguishable from truly Earth-like planets.  Given that simulations of water delivery to habitable zone terrestrial planets predict water mass fractions of $10^{-5}$--$10^{-2}$ \citep{Raymond_2004}, we conclude that most tectonically active planets ---regardless of mass--- will have both oceans and exposed continents, enabling a silicate weathering thermostat.

\begin{acknowledgments}
NBC acknowledges many insightful discussions with J.P.~Townsend, S.D.~Jacobsen, and C.R.~Bina.  The authors also had useful conversations with C.~Andronicos, H.~Gilbert, R.~Jeanloz, M.~Manga, D.~McKenzie, V.S.~Meadows, D.B.~Rowley, J.~Rudge, S.~Stein, D.J.~Stevenson, and R.~Wordsworth. E.S.~Kite provided critical feedback on an early version of the manuscript. The authors thank N.H.~Sleep for sharing an unpublished manuscript.  DSA was supported by an Alfred P.~Sloan research fellowship.
\end{acknowledgments}

\end{document}